\documentclass[sigconf]{acmart}
\AtBeginDocument{%
  }
\newcommand{\circletext}[1]{\raisebox{.5pt}{\textcircled{\raisebox{-.3pt} {\footnotesize{#1}}}}}

\copyrightyear{2026}
\acmYear{2026}
\setcopyright{cc}
\setcctype{by}
\acmConference[CHI EA '26]{Extended Abstracts of the 2026 CHI Conference on Human Factors in Computing Systems}{April 13--17, 2026}{Barcelona, Spain}
\acmBooktitle{Extended Abstracts of the 2026 CHI Conference on Human Factors in Computing Systems (CHI EA '26), April 13--17, 2026, Barcelona, Spain}
\acmDOI{10.1145/3772363.3799003}
\acmISBN{979-8-4007-2281-3/2026/04}

\begin{document}

\title{How Professional Visual Artists are Negotiating Generative AI in the Workplace}


\author{Harry H. Jiang}
\email{hhj@andrew.cmu.edu}
\affiliation{
\institution{Carnegie Mellon University}
\city{Pittsburgh}
\state{PA}
\country{USA}
}

\author{Jordan Taylor}
\email{jordant@andrew.cmu.edu}
\affiliation{
\institution{Carnegie Mellon University}
\city{Pittsburgh}
\state{PA}
\country{USA}
}

\author{William Agnew}
\email{wagnew@andrew.cmu.edu}
\affiliation{
\institution{Carnegie Mellon University}
\city{Pittsburgh}
\state{PA}
\country{USA}
}

\renewcommand{\shortauthors}{Jiang, et al.}

\begin{abstract}

  % Artists have reported that generative AI is impacting them in many ways, including reduced income and appropriation of artistic style. 
Generative AI has been heavily critiqued by artists in both popular media and HCI scholarship. However, more work is needed to understand the impacts of generative AI on professional artists' workplaces and careers. In this paper, we conduct a survey of \textit{378 verified professional visual artists} about how generative AI has impacted their careers and workplaces. We find (1) most visual artists are strongly opposed to using generative AI (text or visual) and negotiate their inclusion in the workplace through a variety of \textit{refusal} strategies (2) there exist a range of factors in artists environments shaping their use of generative AI, including pressure from clients, bosses, and peers and (3) visual artists report overwhelmingly negative impacts of generative AI on their workplaces, leading to added stress and reduced job opportunities. In light of these findings, we encourage HCI researchers to contend more deeply with artists' desires not to use generative AI in the workplace.

\end{abstract}

\maketitle

\section{Introduction}

AI generated art has upended artistic communities in the past few years. Artists have reported economic losses, reputational damage, plagiarism and copyright infringement from AI art being used in place of their labor or to copy their styles \cite{jiang2023ai}.  These findings have been supported by recent studies. \citet{lovato2024foregrounding} surveyed artists at large and found that many think generative is a threat to artists and that generative AI use should be disclosed.

To understand the impacts of generative AI on artistic labor, we center and explore the experiences and perspectives of \textbf{verified professional visual artists}. By analyzing social media posts, \citet{kawakami2024impact} found artists discussing lower income and employment opportunities and increased stress due to generative AI, along with increased distrust towards artists due to suspicion their art was AI generated. A survey of managers and leaders in arts industries found that many firms project that AI will reduce their labor needs \cite{artists_survey}. A survey of the Latin American publishing industry found varying levels of resistance to adoption of generative AI, with workers being more resistant than bosses \cite{genai_writer}. Prior work on the labor impacts of generative AI on artists often focuses on ``artists'' as a general category. However, artists work in a variety of mediums. We build on this prior work by highlighting the workplace experiences of \textit{professional visual artists}.

Through a survey of 378 verified professional visual artists, we found that (1) most participants are strongly opposed to using generative AI (text or visual) and engage in a variety of \textit{refusal} strategies (2) there exist a range of factors in artists' environments shaping their use of generative AI, including pressure from clients, bosses, and peers and (3) participants report overwhelmingly negative impacts of generative AI on their workplaces, leading to added stress and reduced job opportunities. In light of these findings, we discuss how HCI researchers can better support professional visual artists by facilitating non-use \cite{cha2025understanding_nonuse}, resistance \cite{vincent2021data, wong2021tactics}, and unmaking \cite{sabie2023unmaking} in the workplace.

% and identify opportuties for future HCI research

% [Note: Just have a slightly longer introduction as a background]

% \section{Background}

\section{Methodology}

\begin{figure*}[t]
  \centering
  \includegraphics[width=\linewidth]{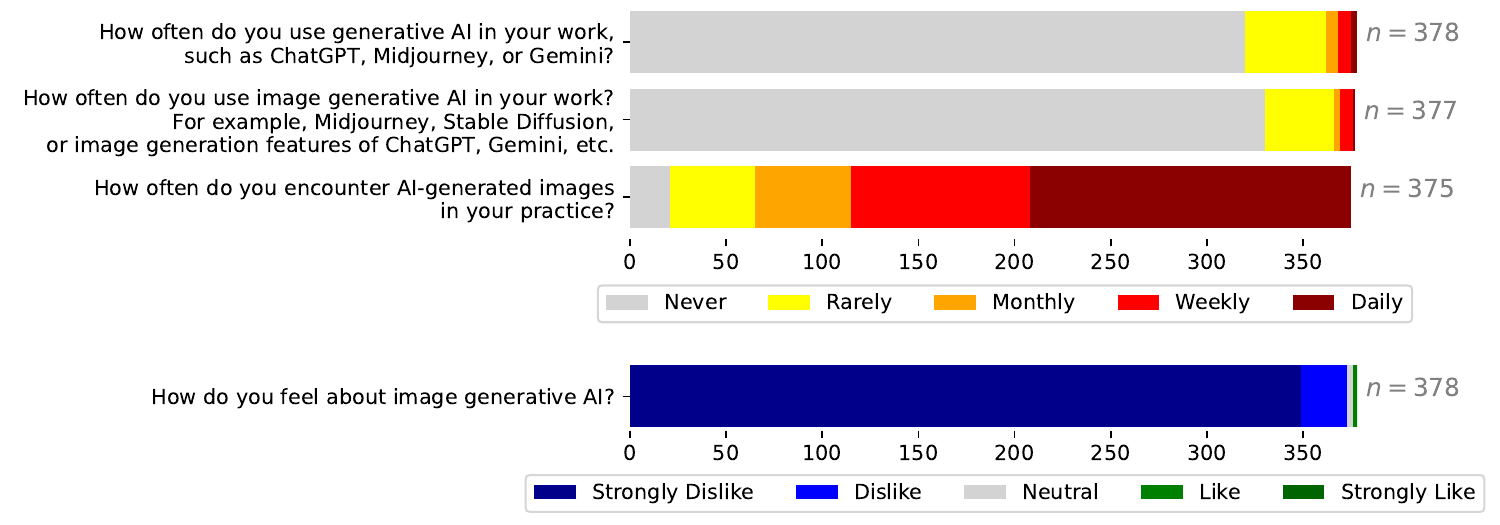}
  \caption{Multiple-choice responses from respondents on the use of, exposure to, and attitude towards generative AI. 85\% of respondents never use generative AI in their work, whereas 88\% never use \textit{image} generative AI. 45\% of respondents encounter AI-generated images in their practice daily, while 25\% do weekly, and 6\% never encounter it. The vast majority of respondents dislike generative AI (99\%), with 92\% expressing a strong dislike.}
  \label{fig:non_use}
\end{figure*}

In this work, we surveyed 378 professional visual artists, seeking to understand \textit{how visual artists feel that generative AI is impacting their careers}. The survey mainly consisted of Likert scale questions with an unstructured text answer field available for respondent to provide details; all questions are optional. The survey is divided in five sections: (1) use of generative AI, (2) general attitudes on genAI, (3) work experiences regarding generative AI, (4) career impacts of genAI, and (5) demographics. We asked demographic questions about income, locality, education, work experience, and various social identities. Our inclusion criteria was that participants must self-identify as ``professional visual artists or have derived any amount of income from the production of visual art.'' The category of ``professional visual artist'' encompasses a wide range of professions, such as illustrators, graphic designers, videographers, and tattoo artists.

\textbf{Positionality.} 
This project is led by a team of researchers which has prior work on HCI studies and projects involving artists %and professional artists 
in their interaction with generative AI. %Such prior work involve interfacing with artists and art community at large with a collaborative approach, and centering artists' views and needs with regards to computing~\cite{jiang2023ai, queer_art_ai}.
This project involves many of the community connections the researchers have built from antecedent studies. All authors are based in computing-area departments at the same university; the first author is also a professional illustrator.

% \subsection{Recruiting and Verifying Professional Visual Artists}

\textbf{Recruitment and Verification.} 
We employed multiple strategies to recruit professional visual artists. The first author shared the survey through flyers at a convention for professional visual artists. Additionally, the first author asked trusted artists to directly share our survey with professional visual artists artists who may be interested. These two methods combined yielded 35 participants; references to them are identified with the symbol \circletext{I} in this text. As these participants were recruited from either a professional conference or trusted artists, we performed no additional steps to verify that these respondents were professional artists. Our remaining 343 participants were recruited through online communities, and identified with the symbol \circletext{V}. Specifically, the first author invited artists in their social network to share an initial screener form on social media, such as Bluesky, Twitter, Tumblr, Discord and Instagram. To mitigate the risk of participant fraud \cite{panicker2024participant_fraud}, this screener asked prospective participants to provide evidence that they are professional visual artists, such as sharing a portfolio. Verified participants were then invited to fill out the survey through a personalized link. We recruited participants between October 2025 and January 2026. Participants received no compensation for completing the survey. However, at the end of the survey, participants were given the opportunity to express interest in participating in a paid interview study .

\textbf{Participant Description.} 
As described above, all demographic questions were optional. Three quarters of participants (279 of 372) report primarily making a living through art. Of the participants who said where they reside (372), the top locations were the United States of America (218), Canada (38), Germany (15), the United Kingdom (14), and France (11). Remaining participants lived in other European countries (40), Latin America (19), Oceania (4), South East Asia (7), or East Asia (3). One participant each lived in Israel, South Africa, and Trinidad \& Tobago. Most participants (279 of 372) identified as primarily making a living through their art. About two thirds (243 of 367) identified as members of the LGBTQ+ community, 19\% (69 of 364) identified as transgender, and 25\% (90 of 367) identified as disabled.

\textbf{Data Analysis.} 
In addition to the descriptive quantitative analyses shared below, we qualitatively examined open-text explanations accompanying questions related to the findings sections described below. To analyze these open-text fields, the three authors conducted collaborative affinity diagramming in a digital whiteboard program \cite{afinity_diagramming}. This allowed us to examine not only whether participants use generative AI, for instance, but also participants' underlying motivations for their (non) use.

\section{Results and Analysis}

\begin{figure*}[h]
  \centering
  \includegraphics[width=\linewidth]{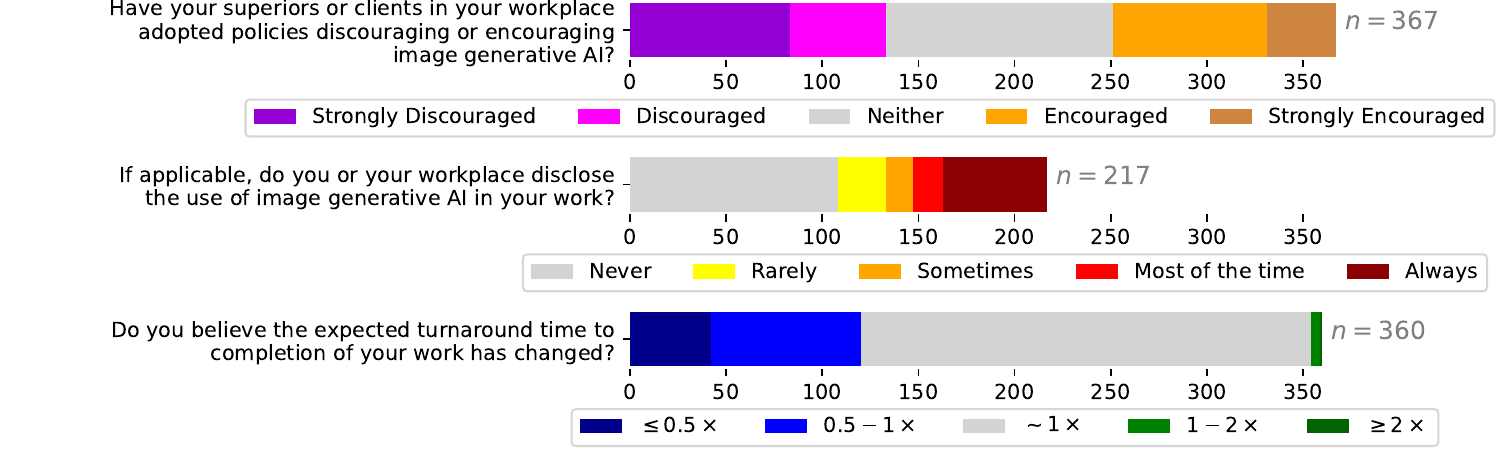}
  \caption{Multiple-choice responses on workplace attitudes and changes regarding generative AI. 32\% of respondents reported neither encouragement or discouragement from superiors or clients, and remaining responses are about evenly distributed between encouragement and discouragement. A majority of respondents (65\%) have not experienced changes in turnaround time of tasks, though almost all of those who reported otherwise have seen shorter expected turnaround times (33\%).}
  \label{fig:work_env}
\end{figure*}

\subsection{How Do Visual Artists \\ (Not) Use Generative AI}

Most participants strongly disliked genAI and chose not to use either text or image models in their practice (Figure~\ref{fig:non_use}). Some initially explored using genAI when models were first released in the early 2020s, but chose to stop using them either because the models did not perform well enough for their purposes or out of growing awareness of ethical concerns related to using genAI. P311 (\circletext{V}; illustrator, designer)``\textit{attempted to use some of these tools to assist in still image and short animation creation, but they have consistently failed to produce useable results.}.''  Likewise, P154 (\circletext{V}; illustrator, painter, designer) said: ``\textit{I experimented with NightCafe and Midjourney when they were being developed (in beta). When I learned they were sourcing their art without permission of the artist, I stopped using them.}''  Among the participants who had used genAI, they often did so because they felt pressured by their employer or a client they engaged with used genAI (more in Section~\ref{sec:work_env}).

Other participants distinguished between using text versus image models. For example, some participants used text based models to help with peripheral tasks related to their art, %like sending emails or copyediting, but chose not to use image models,
such as ``\textit{Chat GPT to help find suitable Amazon marketing keywords,}'' (P299 \circletext{V}; designer, illustrator) or ``\textit{to generate transcriptions of meetings i attend}'' (P318 \circletext{V}; 3D artist). Only a handful of participants actively chose to use visual genAI in their practices, e.g. ``\textit{[Adobe] firefly to make some [...] art with a stock image}'' (P69 \circletext{V}; graphic designer) and ``\textit{Photoshop Creative Suite's built-in generative tools (e.g., generative crop/expand, generative fills, etc.)}'' (P307 \circletext{V}; apparel designer, illustrator). %typically for creating reference images or brainstorming. 
Participants struggled against the increasing integration of genAI into their existing technologies, such as Adobe Photoshop and search engines: ``\textit{I haven't updated my Photoshop in ages, and I keep having to turn off new features in my software.}'' (P250 \circletext{V}; graphic designer, illustrator)

While participants may not have wanted to use genAI in their practice, they frequently encountered AI-generated media in their practice (Figure~\ref{fig:non_use}). For example, artists often reported that they do not use genAI themselves but are sometimes expected to use clients' AI-generated images: ``\textit{I personally don't use image genAI, but I have received generated AI images by clients that I had to springboard off.}'' (P245 \circletext{V}; 3D artist) Participants' AI (non) use is not a one-time choice but rather an ongoing negotiation, especially under the influence of clients and employers. 

\subsection{How Are Visual Artists' Work Environments (Not) Encouraging Generative AI}\label{sec:work_env}

\begin{figure*}[h]
  \centering
  \includegraphics[width=\linewidth]{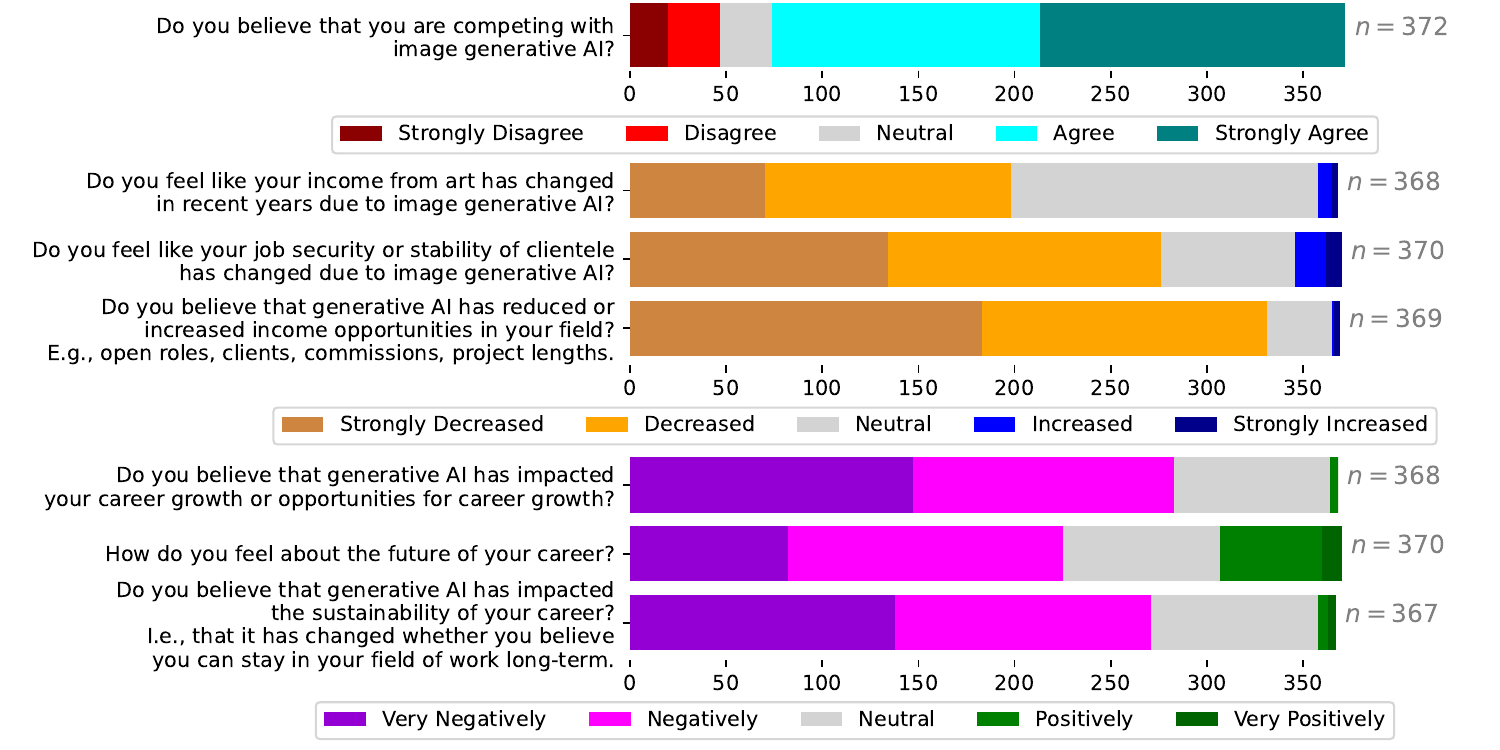}
  \caption{Multiple-choice responses on career impacts of generative AI. The vast majority of artists believe they compete with genAI (80\%). Artists also generally believe genAI has diminished many aspects of their career, such as income (54\%; neutral 43\%), job security or clientele stability (75\%), and income opportunities (90\%). Responses were overall negative regarding career growth (77\%; 1\% positive), future of career (61\%; 17\% positive), and career sustainability (74\%; 2\% positive).}
  \label{fig:income}
\end{figure*}

Our participants' work environments shaped their (non) use of genAI. In multiple-choice responses, those who report discouragement, encouragement, and neither from their superiors or clients are distributed roughly evenly in thirds (Figure~\ref{fig:work_env}). However, in text responses, some participants used the "neither" field to represent mixed messages from different employers. For example, ``\textit{My clients in publishing definitely discourage AI and strongly enforce against using the technology. However, in advertising it is less so; we see it increase through ads in subways, video, etc.}'' (P35 \circletext{I}; illustrator) AI (non) use was also shaped by peer influences: ``\textit{Being caught with AI art is a death sentence as an illustrator in any serious field. Art Directors will blacklist immediately}'' (P228 \circletext{V}; scientific illustrator, concept artist), and ``\textit{Our [tattoo] shop tends to be pretty vocally opposed, but not everyone in the shop is}'' (P319 \circletext{V}; tattoo artist, illustrator, 3D artist).

Among those who said their work discouraged genAI, most artists were freelancers or had client-facing work. Participants reported that thier clients typically discouraged genAI use due to concerns over copyright, quality, and reputation: ``\textit{Many clients have discouraged it, particularly in creative fields that value human work, but also in some fields where copyright, data protection, and NDAs are a concern, where we aren't permitted to use it out of the safety of the company or the integrity of the data/work/research.}'' (P133 \circletext{V}; graphic designer, motion graphic designer, illustrator) In fact, some artists purposefully select clientele who discourage genAI use: ``\textit{I have a somewhat curated clientele, and many of them work with me in part because I do not use generative AI.}'' (P333 \circletext{V}; illustrator, comics artist) This selectivity extended to tools artists use, such as a participant who ``\textit{dropped Adobe [tools] entirely}'' (P45 \circletext{V}; illustrator) due to fears their client's work would be used to train genAI models. For artists who primarily cater to a retail clientele (e.g., convention vendors, shop owners, private commissions, etc.), many report that customers and event organizers often prioritize human creativity and reject genAI: ``\textit{Most who purchase from independent artists do not tend to engage with AI and try not to support those who use it.}'' (P33 \circletext{I}; digital artist, craftsperson), and ``\textit{Most conventions will state that AI is banned in their artist alleys.}'' (P198 \circletext{V}; digital artist)

Among those who said their superiors encouraged AI use, the artists tended to work in more formal employment settings. These employers typically encouraged genAI use due to perceived efficiency gains. For instance, ``\textit{Within my previous employment at a large corporation, AI was heavily pushed as more efficient route and we were often pressured and suggested to use it in our workflows.}'' (P215 \circletext{V}; UI designer, illustrator) Artists often felt coerced into using genAI by their superiors or upper management. One artist reports that ``\textit{My old boss has said that `employees who want a future in this company need to embrace Ai' which is frustrating when every department that is trying to use Ai as directed has reported it does some things fine but other things much worse but we were still told to use it.}'' (P219 \circletext{V}; media and marketing) and another notes that ``\textit{[genAI] is highly encouraged to the point that it feels mandatory}''. (P325 \circletext{V}; animator, illustrator) While our participants often tried to resist genAI by not using the technology, doing so can require negotiating intense internal pressures to use genAI.

\subsection{How Do Visual Artists Feel that Generative AI Has Impacted Their Career}

Visual artists report a strongly negative sentiment towards how AI has impacted their employment and career. Multiple-choice responses on income and career prospect skew heavily negatively on all factors (Figure~\ref{fig:income}). Many report reduced career prospects: ``\textit{They fired me, so it has been a very hard pill to swallow.}'' (P270 \circletext{V}; art director, concept artist, 3D artist) and ``\textit{I’m working on getting out of the field and planning to get my PhD in something non-art related because I can’t see my current work as being sustainable when I see them actively replacing me [with] chatGPT''}. (P214 \circletext{V}; costume designer, illustrator) Demoralization, disempowerment, disrespect, stress, and fear are also commonly expressed, not only regarding individual careers but also extending towards the field at large: ``\textit{It's been pretty demoralizing at times seeing a lot of younger artists giving up because they don't see a future in art. That they're abandoning their creative passions because of AI.}'' (P40 \circletext{V}; illustrator) Some also lamented the public's lack of appreciation of art in their adoption and consumption of genAI and its outputs: ``\textit{It has been demoralizing largely because generated a.i. images look like crap but there is a segment of the population who seem to not care.}'' (P41 \circletext{V}; comic artist and writer, painter)

Of those who felt positively about the future of their career (Figure~\ref{fig:income}), this optimism typicality stemmed from one of three factors (1) they believe in the ongoing or revived appreciation of human-made art by customers, (2) genAI's impact in their field is transient, or (3) they are flexible as a professional to adapt to the disruption. For instance, P83 (\circletext{V}; illustrator, traditional printer) felt that the spread of AI-generated art ``\textit{just brought me more customers and has shown me, that people do care about handmade things much more than one thinks!}'' Many do note that they feel lucky to be in such positions to persist: ``\textit{I now survive on a single freelance contract that is so hyper specific it is difficult to replace. This is due to happenstance and otherwise I would be completely out of luck.}'' (P37 \circletext{V}; graphic designer, illustrator)

Of those who felt negatively, some visual artists worried about the burden of being asked to prove that their artwork \textit{is not} AI-generated. P124 (\circletext{V}; illustrator, designer) said, ``\textit{I find users online to be more critical, looking at art less to enjoy it and more so to figure out if its AI generated or not. There's a lot of pressure and anxiety in proving you are a real person now.}'' Likewise, P79 (\circletext{V}; digital artist, traditional artist, sculptor) expressed: ``\textit{I have seen false accusations for use of AI in work from other artists who do not use AI and I am fearful of being accused of this as well, I now record the creation process of most things so that I have proof AI was not utilized.}''

\begin{figure*}[h]
  \centering
  \includegraphics[width=\linewidth]{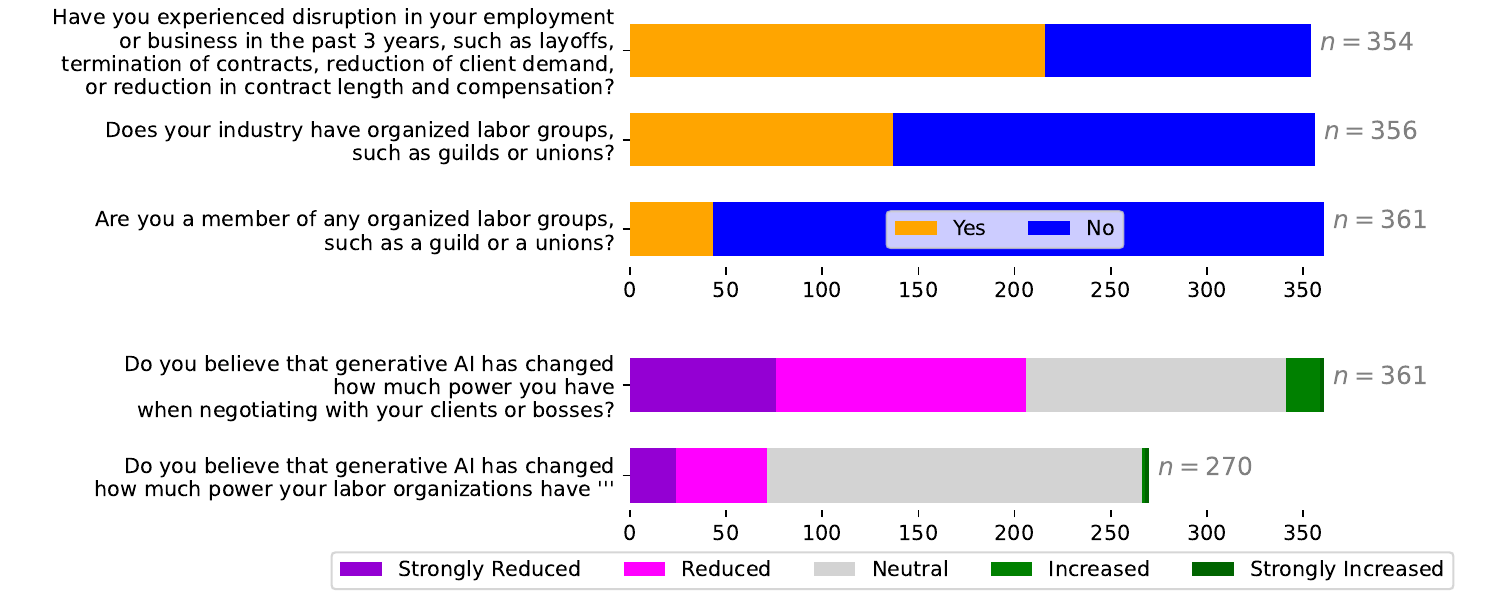}
  \caption{Multiple-choice responses on labor disruption and organized labor. 61\% of respondents report disruption in employment or business. 38\% of artists ($n=137$) work in an industry with organized labor, but only 12\% are members ($n=43$). Some respondents perceive diminished negotiation power, with 57\% expressing reduction of individual's power (6\% for increase) and 26\% ($n=71$) for reduction of labor organizations' power (1\%, $n=4$ for increase), though note that the latter question includes responses from artists who do not work in fields with a labor organization.}
  \label{fig:unions}
\end{figure*}

As the pressure to use genAI often comes from superiors in the workplace (Section~\ref{sec:work_env}), there were also many reports of tension between artists and bosses over genAI. P34 (\circletext{I}; illustrator, animator, designer) said: ``\textit{It has created tense situations between me and my managers. We do not see eye to eye at all, and often dance around the topic to avoid tensions.}'' Additionally, P165 (\circletext{V}; art director) felt that``\textit{only the CEO and staff immediately beneath him seem to like [genAI].}'' On matters of negotiation, artists overall report decreased power with respect to their clients and bosses, and some believe labor organizations' power are also decreased (Figure~\ref{fig:unions}).

\section{Limitations}

% [Something about limitations of recruitment]
A major limitation of our work is that most of our participants reside in the United States (218), Europe (80), or Canada (38). This could be attributable to the social network of the authorship based in a US university, as well as our survey being written in English. Given the global nature of artistic labor, there is a need for additional research on the impact of generative AI on artists in the Global Majority \cite{bengali_art_ai_impact, non_western_art}, both in domestic artistic productions and global supply chains of artistic labor \cite{Yoon03092017, outsourcing}. We also acknowledge the possibility of sampling bias through recruitment strategies. The survey is presented to participants as seeking to study impacts on the workplace, which can cause over-representation of stronger and negative responses and the inverse for those with neutral or positive stances. In addition to surveys in other languages, we invite other modalities of research on this topic, such as ethnography and interview methods. We do not claim that our findings represent the experiences of all professional visual artists.

% Furthermore, since our survey is shared through peer networks and online social networks, the sharing behavior of artists similarly affects participation of more vocal respondents.

\section{Conclusion and Future Work}

% [Point on textual models feeling more acceptable visual models]
% [Employers are not a monolith, some discourage some encourage]
In light of participants' negative attitudes toward genAI, they tried to \textbf{abstain from using genAI} as a strategy of resistance. At the same time, participants felt increasingly pressured to use genAI by both their employers and the technology companies that make their creative tools. From a labor and economic perspective, genAI disrupts the relation between artists and their employers and clients. Artists perceive a more constrained and unpredictable labor market and increased pressure at work and in their careers. Even though we find that employers and clients actually have a high diversity of attitudes towards genAI, in cases where the artist and the employer or client is misaligned, artists feel increased pressure to use genAI, and are in some cases forced to use it against their wishes. These build on understandings on AI non-use and refusal in other professional settings \cite{cha2025understanding_nonuse, wells2025resistance}, motivating further development of tools to facilitate non-use  and resistance. Specifically, while many technical methods to disrupt AI systems exist, such as anti-facial recognition tools \cite{shan2020fawkes, cherepanova2021lowkey}, Glaze \cite{shan2023glaze}, and Nightshade \cite{shan2024nightshade}, acceptance, usage patterns, and adaptations to user needs remain understudied \cite{logas2024subversive}.

% [Marginalized communities in the workforce]
% As a result, 
Participants felt reduced progressional agency, impacting their ability to make a living as artists. The negative impact of generative AI on visual artists may disproportionately harm members of marginalized communities. As described above, almost two thirds of participants identified as LGBTQ+, one quarter identified as disabled, and one fifth identified as transgender. As art making often takes place outside the boundaries of in-person salaried employment, art may be a more accessible career for people with disabilities or those who experience employment discrimination, like LGBTQ+ people. From that perspective, the ingress of image genAI into the visual arts economy may further harm these marginalized communities. While recent work has explored how LGBTQ+ \cite{queer_art_ai} and disabled \cite{disabled_artists_ai} artists negotiate genAI use, those most critical of genAI may be unlikely to participate in a study on genAI use. Therefore, we call for additional research focused specifically on LGBTQ+ and disabled artists who actively choose not to use or to resist genAI.

% Computer scientists are increasingly claiming that AI tools will  on systems Amid the increasing calls for HCI researchers to design tools to ``augment'' the work of art artists 
Amid the increasingly dubious rhetoric that AI will help "augment" workers \cite{augmentation_washing_2025}, our findings demonstrate the need to consider whether workers want to be "augmented" in the first place. Rather than trying to design "better" genAI models for artists, we encourage HCI researchers to contend more deeply with genAI non-use. 

% non-use and is a need to focos on the experinces of 
% or LGBTQ+ people.
% As art making often takes place outside the bounds of salaried employment, art making 

% \begin{acks}
% \end{acks}

\bibliographystyle{ACM-Reference-Format}
\bibliography{sample-base}

\appendix

\end{document}